\documentstyle[12pt,geompsfi]{article}
\input epsf
\newcommand{\nc}{\newcommand}

\newcommand{\bea}{\begin{eqnarray}}
\newcommand{\eea}{\end{eqnarray}}
\newcommand{\be}{\begin{equation}}
\newcommand{\ee}{\end{equation}}
\newcommand{\bc}{\begin{center}}
\newcommand{\ec}{\end{center}}
\newcommand{\ba}{\begin{array}}
\newcommand{\ea}{\end{array}}
\newcommand{\btab}{\begin{tabular}}
\newcommand{\etab}{\end{tabular}}
\newcommand{\bfig}{\begin{figure}}
\newcommand{\efig}{\end{figure}}
\newcommand{\vek}[1]{{\rm\bf #1}}
\newcommand{\non}{\nonumber}

\newcommand{\Psibar}{\overline{\Psi}}
\newcommand{\ubar}{\overline{u}}
\newcommand{\eps}{\varepsilon}

\newcommand{\sslask}{\!\!\!/}
\nc{\dbar}{d \hspace{-0.25em} \raisebox{0.45ex}{\Large -}}
\nc{\ra}{\rangle}
\nc{\la}{\langle}

\newcommand{\eq}[1]{Eq.(\ref{#1})}

\newcommand{\refc}[1]{\raisebox{1ex}{\scriptsize \ref{#1}}}
\nc{\refs}[2]{\raisebox{1ex}{\scriptsize \ref{#1}, \ref{#2}}}
\nc{\fig}[1]{\mbox{Fig.~\ref{#1}}}

\nc{\cO}{{\cal O}}
\nc{\bort}[1]{}

\def\Im{{\rm Im}\hskip2pt}
\newcommand{\figcap}[1]{\refstepcounter{figure}
	{\bf Figure \thefigure}: {\small #1}}

\newcommand{\prl}[3]{{\it  Phys.~Rev.~Lett.~} {{\bf #1} {(#2)} {#3}}}
\newcommand{\pr}[3]{{\it  Phys.~Rev.~} {{\bf #1} {(#2)} {#3}}}

\nc{\np}[3]{{\it  Nucl.~Phys.~ }{{\bf #1} {(#2)} {#3}}}
\nc{\app}[3]{{\it Astropart.\ Phys.\ }{{\bf #1} {(#2)} {#3}}}
\nc{\si}{\hat\Sigma}
\nc{\ve}{\varepsilon}
\nc{\amm}{anomalous magnetic moment}
\nc{\dega}{\Delta E^{\gamma}}
\nc{\dee}{\Delta E^{e^+e^-}}
\nc{\dele}{\Delta E}

\nc{\dm}{\Delta m}
\nc{\dmb}{\delta  {\cal M}_B}
\nc{\mhz}{{\hat{\mu}_z}}
\nc{\om}{\omega}
\begin{document}
\vspace*{-10mm}
\begin{flushright}G\"oteborg ITP 95-30\\ hep-ph/\\
	 October 1995\\[-5mm]\end{flushright}
\begin{center}
{\bf ELECTRON THERMAL SELF-ENERGY IN A MAGNETIC FIELD\footnote[2]{
	Contribution to {\sl The 4th International Workshop on Thermal
	Field Theories and Their Applications}, Dalian, China,
	August 6-12, 1995.}}\\[15pt]
DAVID PERSSON\\
\baselineskip 13pt
\begin{it}
   	Chalmers University of Technology and
	G\"oteborg University,\\ \baselineskip 13pt
 Institute of Theoretical Physics, S-412 96 G\"oteborg,
	Sweden\\ \baselineskip 13pt
\end{it}
Email: tfedp@fy.chalmers.se\\[10ex]
\end{center}
\begin{center}
\leavevmode
\epsfbox{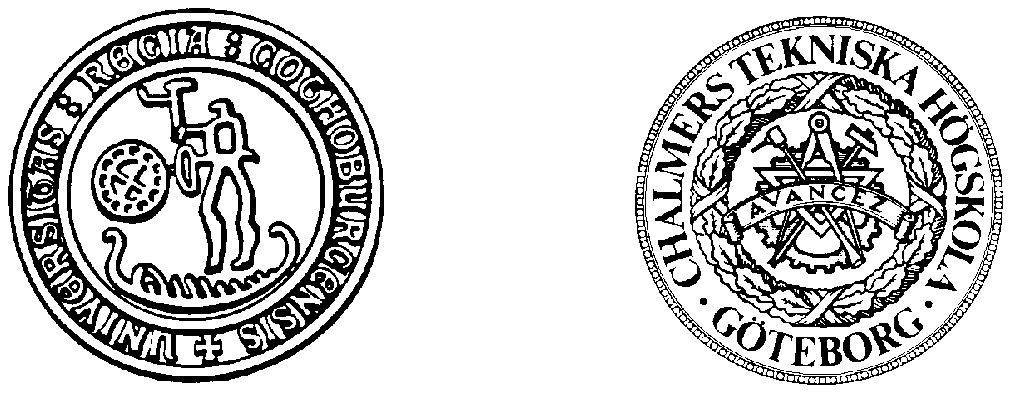}\vspace{1ex}
\end{center}
\begin{center}{ABSTRACT}\end{center}
\begin{quote}
Using the general form of the static energy solutions to the Dirac equation
with a magnetic field, we calculate a general self-energy matrix in the
Furry-picture.
 In the limit of high temperatures, but even higher magnetic fields,
a self-consistent dispersion relation is solved.
In contrast to the high temperature limit,
this merely results in a small mass
shift. The electron anomalous magnetic moment is calculated. The contribution
from thermal fermions is found to be different from the corresponding
contribution using perturbation theory and plane-wave external states.
In the low temperature limit the self-energy is shown to exhibit
de Haas--van Alphen oscillations. In the limit of low temperatures and high
densities, the self-energy becomes very large.
\end{quote}
\newpage
%
\section{\bf Introduction}
Several compact stellar objects (e.g. neutron stars, white dwarfs and red
giants) and cosmological models are characterized by high temperatures and/or
densities, and also the presence of a strong background magnetic field.
It is thus of great interest to compute the properties of electrons in
such extreme environments. Earlier the effective potential has been
considered\refc{elmforsps94}.
Here we shall deal with the electron self-energy, important for kinematical
issues. Apart from the general self-energy matrix in the Furry-picture, and
the treatment of the de Haas--van Alphen oscillations in the self-energy, this
work is merely a shorter version of results published
elsewhere\refc{elmforsps95}, to which we refer for further details and
references.
\section{ \bf Furry--Picture Propagator}
The Dirac equation for
an electron (charge -e) in an external field, with the vector potential
$A^\nu$, reads
\be
	(i \partial\sslask +e A\sslask -m )\Psi_\kappa^{(\pm)}(x)=0~~~,
\ee
where $\kappa$ denotes a complete set of quantum numbers necessary to
specify the solutions, and the superscripts $+$ and $-$
 are referring to positive and negative
energy solutions, respectively.
Using a complete set of static energy solutions we may represent the second
quantized electron field in the Furry-picture\refc{furry}
\be
	\Psi(\vek{x},t)=\sum_\kappa\left[b_\kappa\Psi^{(+)}_\kappa(\vek{x},t)+
    d^\dagger_\kappa\Psi^{(-)}_\kappa(\vek{x},t)\right]\ ,
\ee
where  $(b,d)$ are the standard annihilation operators for particles and
anti-particles.
The fermion  propagator, including the effects of some distribution of
particles, can then be constructed explicitly as the expectation
value\refc{elmforsps94}
\be
   iS(x',x)=\la \bf T \left[\Psi(x')\Psibar(x)\right] \ra~~~.
\ee

In the case of a static uniform magnetic field in the negative $z$-direction
we may choose $A_\nu=(0,0,Bx,0)$. The solutions to the Dirac equation
must then be of the form\refc{elmforsps95}
\be
	\label{wavefun}
	\Psi^{(\pm)}_{\zeta,n,p_y,p_z}(\vek{x},t)=\frac1{N_\kappa}
	\exp[\pm i(-E_n t +p_y y + p_z z)] V_{n,p_y}(x)
	u_{\zeta,n,p_y,p_z}~~~,
\ee
where $\zeta=\pm 1$ is a polarization index, $n =0,1,2, \ldots$,
and $p_y,p_z \in {\bf R}$. The energy spectrum is given by
\be
	E_n(B,p_z)=\sqrt{m^2+p_z^2+2eBn}~~~.
\ee
In the chiral representation of the $\gamma$-matrices we have
\be
	 V_{n,p_y}(x)={\rm diag}[I_{n,p_y}(x),I_{n-1,p_y}(x),
	I_{n,p_y}(x),I_{n-1,p_y}(x)]~~~,
\ee
where $I_{n,p_y}(x)=(eB/\pi)^{1/4}\exp[-eB/2\, (x-p_y/eB)^2]1/\sqrt{n!}\,
	H_n[\sqrt{2eB}(x-p_y/eB)]$. Here $H_n$ is a Hermite polynomial,
and we define $I_{-1,p_y}=0$.
The spinor $u_{\zeta,n,p_y,p_z}$ is independent of $x^\nu$. There is
a twofold degeneracy corresponding to $\zeta=\pm 1$ in all but the lowest
Landau level $n=0$. The electron propagator is with this choice of gauge
\bea
	\lefteqn{S(x'^\nu,x^\nu)=\sum_{n=0}^\infty
	\int \dbar k_0 \dbar k_y \dbar k_z\,
	S_{n,k_0,k_y,k_z}(x',x)} \non \\
	&& \times \left[ \frac1{k_0^2-m^2-k_z^2-2eBn+ i \eps} + 2 \pi i
	\delta(k_0^2-m^2-k_z^2-2eBn)f_F(k_0) \right]\non \\
	&& \times \exp[-i k_0(t'-t)+i k_y (y'-y) + i k_z (z'-z)]~~~,
\label{elecprop}
\eea
where $\dbar^n k \equiv d^n k/(2\pi)^n$, and $f_F(k_0)\equiv
\Theta(k_0)f_F^+(k_0)+
 \Theta(-k_0)f_F^-(k_0)$.  In the case of thermal equilibrium the one-particle
 fermionic distribution is
$f_F^{\pm}(k_0)= 1/(e^{\beta(k_0 \mp \mu)}+1)$,
where $\beta$ is the inverse temperature and $\mu$ is the chemical
potential determined by the charge density of electrons and
positrons of the system.
Not specifying the representation of the $\gamma$-matrices
 we may write\refc{mak95}
\bea
S_{n,k_0,k_y,k_z}(x',x)&=& (k_0 \gamma_0-k_z \gamma_z+m) [\sigma_+
	I_{n,k_y}(x')I_{n,k_y}(x) +\sigma_- I_{n-1,k_y}(x')I_{n-1,k_y}(x)]
 \non \\
	&&-i \sqrt{2eBn}[\gamma_+ I_{n,k_y}(x')I_{n-1,k_y}(x)-
	\gamma_- I_{n-1,k_y}(x')I_{n,k_y}(x)]~~~,
\eea
where $\gamma_\pm \equiv (\gamma_x \pm i \gamma_y)/2$, and $\sigma_\pm \equiv
(1\pm \sigma_z)/2~~( \otimes {\bf 1} \mbox{\rm when necessary})$.
\section{\bf The General Self-energy Matrix}
There are two possible contributions to the one-loop electron self-energy.
The tadpole contribution is proportional to the total electric charge and
current in the medium\refc{elmforsps95},
 and thus vanishing in a neutral environment.
Therefore we are left only with the 1PI self-energy
\be
	-i\Sigma(x',x)=(-ie)^2\gamma_\mu iD^{\mu\nu}(x'-x)iS(x',x)
	\gamma_\nu~~~.
\ee
The photon propagator may be written
\be
	iD^{\mu\nu}(x)=\int \dbar^4 q e^{-i q\cdot x}  \left( g^{\mu\nu}-
        \xi\, q^\mu q^\nu \frac{\partial}{\partial q^2}\right)  \left[
        \frac{-i}{q^2+i\ve} - 2\pi \delta(q^2)f_B(q_0) \right]~~~,
\ee
where the photon distribution function in the case of thermal equilibrium is
$f_B(q_0)=1/(\exp[\beta |q_0|]-1)$.
The effective Dirac equation may then be written
\bea
\lefteqn{\int d^4 x \Psibar^{(+)}_{\zeta,n,p_y,p_z}(x) (i \partial\sslask
	+e A\sslask -m) \Psi^{(+)}_{\zeta',n,p_y,p_z}(x)=} \non \\
	&=& \int d^4 x
	\int d^4 x' \,  \Psibar^{(+)}_{\zeta,n,p_y,p_z}(x) \Sigma(x,x')
	\Psi^{(+)}_{\zeta',n,p_y,p_z}(x') \non \\
&\equiv & \frac1{N_\kappa^2} \int dt\, dy \, dz \,  \ubar_{\zeta,N,p_y,p_z}
\hat\Sigma_N(B,p_z;p_y) u_{\zeta',N,p_y,p_z}~~~,
\label{effdirac}
\eea
where we have used the general form of the electron wave-functions in
\eq{wavefun} to define a general self-energy matrix $\hat\Sigma_N$.
The space-time integral (i.e. energy---momentum conservation) implies that both
 sides of \eq{effdirac} are diagonal in $n,p_y$ and $p_z$.
On-shell, i.e. with tree-level wave-functions, the effective Dirac
equation~(\ref{effdirac}) results in an energy shift
$E_{\zeta,n}^{(1)}(B,p_z)=E_n(B,p_z)+ \dele_{\zeta,n}(B,p_z)$, where
\be
\label{energyshift}
	\dele_{\zeta,n}(B,p_z)=\ubar_{\zeta,n,p_y,p_z}
	\hat\Sigma_n(B,p_z;p_y) u_{\zeta',n,p_y,p_z}~~~,
\ee
that actually is independent of $p_y$. The on-shell self-energy has also
been shown\refc{elmforsps95} to be
independent of the gauge-fixing parameter $\xi$.
If not otherwise stated we shall use the Feynman gauge $\xi=0$.
\section{\bf Strong Field Limit}
In favorable conditions the gauge-fixing dependence of the self-energy may
be neglected also off-shell. Then we may move forward and solve  a
self-consistent dispersion relation. It is a well-known fact that in the
high-temperature limit this is the case. Self consistent dispersion-relations
have been solved in the high temperature limit without\refc{weldon89}, and also
with a magnetic field using the Schwinger
 proper time method\refs{elmforsps95}{elmfors95}.
The result, rich in physics,  contains for example particle as well as
 hole solutions.
With the general self-energy
matrix presented here it is possible to perform the same analysis also in
the Furry-picture.

On the other hand the Furry-picture
is particularly suitable in the strong field limit. In this limit the dominant
contribution will come from intermediate electron states in the lowest Landau
 level ($n=0$) only. In the limit $\{eB\gg T^2 \gg m^2,p_z^2,\mu^2\}$
we find in the lowest Landau level, using only $n=0$ in the electron
propagator \eq{elecprop}
\bea
	\label{sig1}
	\si_{0}^{\rm vac} &\simeq& \sigma_+ m \frac\alpha{4\pi} \ln^2\left(
	\frac{2eB}{m^2}+ \Delta-1 \right)~~~,\\
	\label{sig2}
 \si_{0}^{e^+e^-}  &\simeq& -\sigma_+ m \frac\alpha{\pi}\left[
	\ln^2\left( \frac{T}{m} \right) -2 \ln\left( \frac{T}{m} \right)
	\ln\left( \frac{eB}{m^2\sqrt{\Delta+1}} \right) \right]~~~,\\
	\si_0^{\gamma} &\simeq &  -\sigma_+ m \frac{2\alpha}{\pi}
	\ln\left( \frac{T}{m} \right)~~~,
	\label{sig3}
\eea
where we have defined $m^2 \Delta\equiv E^2-m^2-p_z^2$, as the deviation
from the on-shell energy eigenvalue. The self-energy
$\si_0=\si_{0}^{\rm vac}+ \si_{0}^{e^+e^-}+\si_0^{\gamma}$,  has
been split into its contributions from the vacuum, thermal fermions, and
thermal photons, respectively. Similarly the gauge-fixing dependent parts
of the off-shell self-energy are in this limit
\bea
 \si_{0}^{{\rm vac}(\xi)}&\propto & \xi m \alpha \Delta  \ln\left(
	\frac{2eB}{m^2}\right)~~~,\\
	 \si_{0}^{{e^+e^-}(\xi)}&\propto & \xi m \alpha \Delta \left(
\frac{mT}{2eB} \right)^2  \ln\left(
	\frac{2eB}{m^2}\right)~~~, \\
	\si_{0}^{{e^+e^-}(\xi)}&\propto & \xi m \alpha \Delta  \ln\left(
	\frac{T}{m}\right)~~~.
\eea
Keeping only terms $\cO[\ln^2(2eB/m^2)]$ and $\cO[\ln(2eB/m^2)\,
\ln(T/m)]$, we may thus neglect the gauge fixing dependence. Also notice that
all of the three different contributions to the $\xi$ dependent part of
the self-energy are proportional to $\Delta$. We have thus explicitly shown
that the on-shell (i.e. $\Delta=0$) self-energy is gauge-fixing independent
in this limit.
The self-consistent dispersion relation obtained from \eq{effdirac} reads
in this limit
\be
	\label{strongdirac}
	(E\gamma_0 -p_z \gamma_z -m) \sigma_+ u_0 = \dmb
	\sigma_+ u_0~~~,
\ee
where we have defined the ``thermo-magnetic mass''
\be
	\dmb =m \alpha/(4\pi)\,\left[ \ln^2\left(\frac{2eB}{m^2}\right)+
	 8\ln\left(\frac{T}{m}\right)\ln\left(\frac{eB}{m^2}\right)
	-4\ln^2\left(\frac{T}{m}\right)\right]~~~.
\ee
The presence of $\sigma_+$ just signifies that we have already used that
the wave-function in the lowest Landau level is proportional to the matrix
$ V_{0,p_y}(x)=I_{0,p_y}(x)\, {\rm diag}[1,0,1,0]$, according to \eq{wavefun}.
The above \eq{strongdirac} is thus equivalent to the tree-level Dirac
equation with the mass
$M\equiv m+ \dmb $. Unlike in the high temperature limit,
this self-consistent dispersion relation thus only results in an energy
shift
\be
	E_0^{(1)}=\sqrt{(m+ \dmb)^2+p_z^2}~~~.
\ee
To the lowest order in $ \dmb$ this is the same as would have been
obtained from \eq{energyshift}. How large can  then $\dmb$ be?
In order for $\dmb \approx m$ we must have $eB/m^2 \approx 10^{17}$,
i.e. $B\approx 10^{27}T$, an immensely large field. However, this makes our
negligence of the $\xi$ dependent parts more accurate then it appeared at
first when keeping only the leading logarithms. The gauge fixing dependent
part of the self-energy is proportional to $\Delta \approx \dmb
m/E_0$, that is small for most magnetic field strengths.
\section{\bf Weaker Magnetic Fields}
In the case of an arbitrary  magnetic field the sum over all Landau levels
in the electron propagator has to be considered. This problem has recently
been solved\refc{elmforsps95}.
Let us consider an on-shell electron in the lowest Landau level  with
vanishing momentum. We may now perform an expansion to linear order in
$eB/\lambda^2$, where $\lambda$ must be some energy scale, initially not
known in this naive expansion. The result for the  contributions
to the self-energy from vacuum and thermal photons are, respectively,
\bea
	\dele_0^{\rm vac} &\simeq & -\frac{eB}{2m}\frac{\alpha}{2\pi}~~~, \\
 	\dega_0 &\simeq & m \frac{\alpha \pi}{3} \frac{T^2}{m^2} +
	\frac{eB}{2m} \frac{2\pi\alpha}9 \frac{T^2}{m^2}~~~.
\eea
The thermal electron contribution is more involved. In the absence of
a magnetic field we find the well-known result
\bea
      \lefteqn{  \dee_0(B=0,p_z=0) \equiv \dm^{e^+e^-}~~~~~~~~~~~~~~~~}\non \\
      & =& \frac{\alpha}{2\pi} \int_{-\infty}^\infty dk_0
        \Theta(k_0^2-m^2) f_F(k_0) \sqrt{k_0^2-m^2}
        \frac{k_0-2m}{m(k_0-m)}~~~.
\label{deltame}
\eea
To linear order in $eB$ we find\refc{elmforsps95}
\bea
        \dee_0&=&\dm^{e^+e^-} +
        \frac{eB}{2m}\frac{\alpha}{3\pi} \int_m^\infty
        \frac{d\omega}{\sqrt{\omega^2-m^2}} \non \\
        && \times \left[ \left( \frac{2\omega^2+
        2m\omega-m^2}{m^2}-\frac{m}{(\omega+m)} \right)f_F^+(\omega)
         + 2m \frac{df_F^+(\omega)}{d\omega} \right]+ \non \\
        &&  +\frac{eB}{2m}\frac{\alpha}{3\pi} \int_m^\infty
        \frac{d\omega}{\sqrt{\omega^2-m^2}} \frac{ 2\om^3-3m^2\om -2m^3}
        {m^2(\om+m)} f_F^-(\om)~~~,
\label{expvalseel}
\eea
where some integrations by parts has been performed in order to make the
result infra-red finite. The spin energy-shift for a particle of charge $-e$,
with spin $({\bf s})$ magnetic moment $\mbox{\boldmath $\mu$} =-e/(2m)g \la
{\bf s} \ra$
in a magnetic field ${\bf B} =-B \hat z$ is $\dele = -\mbox{\boldmath $\mu$}
\cdot {\bf B}$.
We may thus write
\be
	\dele= \dm^{\beta,\mu}-\frac{eB}{2m} \frac{\delta g}2~~~,
\ee
where $ \dm^{\beta,\mu}$ is the thermal mass, and $\delta g/2$ is the anomaly
of the magnetic moment. Usually $\delta g/2$ is obtained from the vertex
correction (i.e. the triangle diagram). However, in the case of the
contribution from thermal electrons the results differ. The discrepancy is
caused by the external states being Landau levels, and not plane waves, as
assumed when considering the vertex correction. Notice that the anomaly is
obtained from the transverse part of the vertex correction, not included in
 the Ward-identity relating the vertex to the self-energy.

Let us now consider the limit of vanishing temperature, but finite density
of electrons. We may then perform the energy integral in \eq{deltame} and
\eq{expvalseel} to find
\bea
	\dele_0^{e-}&\simeq& \frac{\alpha}{4\pi} \left[ \frac{\mu-2m}{m}
	\sqrt{\mu^2-m^2}-3m\ln\left( \frac{\mu+\sqrt{\mu^2-m^2}}{m} \right)
	\right] \non \\
	 && + \frac{eB}{2m} \frac{\alpha}{3\pi}\frac{\sqrt{\mu^2-m^2}}{m}
	\left[ \frac{\mu}{m} +2-\frac{m}{\mu+m}- \frac{2m^2}{\mu^2-m^2}
	\right]~~~.
\label{setzero}
\eea
This may become very large at high densities, like
for example in the case of a neutron star,
where $\mu/m \approx 10^2$, and $T/m \approx 1$. Comparing with the
high temperature case, we see that the leading terms $m(e\mu/m)^2$ has a
similar origin to the corresponding $m(eT/m)^2$ terms. Thus this contribution
has been obtained from a ``hard dense loop'', and higher loop corrections
will only give sub-dominant  contributions\refc{braatenp90}
 in powers of $(e^2 \mu/m)$.
\section{\bf De Haas---van Alphen Oscillations}
Notice that the self-energy given in \eq{setzero} will become divergent as
$\mu \rightarrow m^+~,~~T=0$. This is not physically acceptable, since
the dense and thermal contribution should vanish in this limit. It seems as if
our naive expansion in this case has been in powers of $eB/(\mu^2-m^2)$,
and thus is not valid for small densities. On the other hand, if $2eB \approx
\mu^2-m^2$ only a few lower  Landau levels will be occupied by electrons.
We thus have a situation very similar to the strong field limit, well
suited for the  explicit spectral decomposition.
Split the thermal electron part  of the self-energy into
its contributions from the different Landau levels:
\be
	\dele_0^{e^-}=\sum_{n=0}^N \dele_0^{e^- (n)}~~~,~N
	={\rm Int}[(\mu^2-m^2)/2eB]~~~.
\ee
Defining  $a_n \equiv m(\om-m)/2eB -n $, and
$\Theta_n=1~,~n\geq 0;~\Theta_n=0~,~n \leq -1$, we find
\bea
	\lefteqn{\dele_0^{e^- (n)}=\frac{\alpha}{\pi}\int_{\sqrt{m^2+
	2eBn}}^\mu
	\frac{d\om}{\sqrt{\om^2-m^2-2eBn}}\left\{
	 \left[ \frac{m}{n!}a_n+\Theta_{n-1} \frac{\om-m}{(n-1)!}\right]
	\right.}\non \\
 	&&\times \left.\left[ (-a_n)^{n-1} e^{a_n}{\rm E}_1(a_n-i\ve) +
	\Theta_{n-2}
	\sum_{l=0}^{n-2} (-a_n)^l (n-2-l)! \right]-  \frac{m}{n!} \Theta_{n-1}
	\right\}~~~.
	\label{selfdhva}
\eea
In the limit $\mu \rightarrow m$ only the lowest Landau level will contribute.
Using the series expansion of the exponential integral, we find for
$\mu^2-m^2 \ll 2eB$
\be
	\dele_0^{e^-} \simeq -m\frac{\alpha}{\pi}
	 \left\{ \ln\left(\frac{eB}{m^2}
	\right)\, \ln\left( \frac{\mu+\sqrt{\mu^2-m^2}}{m} \right) +
	\sqrt{2\frac{\mu-m}{m}} \left[ 2- \ln\left(\frac{\mu-m}{m} \right)
	\right] \right\}~~~.
\ee
We can thus see that the electron contribution to the self-energy is
vanishing as $\mu \rightarrow m^+~,~~T=0$, and furthermore that it is
non-analytical in $B$. As $(\mu^2-m^2)/2eB $ is becoming larger, consecutive
Landau levels will cross the Fermi-surface, and start contributing.
The contribution from the $n$-th Landau level is vanishing as $\mu  \rightarrow
\sqrt{m^2+2eBn}$, so the self-energy is continuous. However, the derivative
with respect to the chemical potential is diverging  as $\mu  \rightarrow
\sqrt{m^2+2eBn}$. These sharp cusps will be smoothed out at finite temperature.
There are thus  oscillations  in the self-energy at low temperatures,
and Fermi-momentum
squared of the order of the magnetic field.
As a function of  $1/eB$, they have a quasi-periodicity of
 $2/(\mu^2-m^2)=2\pi/A$, where $A$ is the area of an extremal cross section
of the Fermi-surface.
Such oscillations are well-known as de Haas--van Alphen oscillations
in the magnetization of a
Fermi-gas, but not earlier found in the fermion self-energy.

For $n\geq 1$, $a_n$ may become negative. This will cause an imaginary part
in the self-energy according to ${\rm E}_1(-x-i\ve)=-{\rm Ei}(x) +i \pi$, for
$x>0$. Let $\Gamma_{e^- \mapsto e^- \gamma}$ denote the total decay rate for an
electron in the Fermi sea decaying into an electron in the lowest Landau level,
that not is supposed to be occupied. Then we find
\be
	\Gamma_{e^- \mapsto e^- \gamma}=2 \Im \dele_0^{e^-} ~~~.
\ee
An ostensible contradiction appears  here.  The lowest Landau level, always
below the Fermi energy, is supposed  not to be occupied.
However, perturbation theory rests on the assumption on
adiabatic turn on/off of the interaction as $t \rightarrow \mp \infty$.
The external state is thus supposed to be separated from the heat and
charge bath before and after the interaction is taking place, and not
affected by the occupation of states in the medium.
In \fig{dhva} we show the real and imaginary  parts of the dense
electron contribution to the self-energy as a function of the inverse magnetic
 field. Unfortunately the cusps
in the real part, as Landau levels with $n \geq 2$ are crossing the
Fermi-surface  cannot be distinguished on this scale.
The cusp in the imaginary part at $\mu=m+2eB/2m$ follows from \eq{selfdhva},
but has no counterpart in ordinary de Haas--van Alphen oscillations.
\begin{figure}[tbp]
\centerline{ \psfig{figure=dhvaplot,height=8cm}}
\vspace{6ex}
\figcap{The  real part and the imaginary part (dotted line) of the electron
self-energy exhibits de Haas--van Alphen
 oscillations at low temperatures. Here $T=0$ and $\mu=\sqrt{3}m$.
 Notice the scaling
of the real part, that actually is negative. }
  \nopagebreak
  \label{dhva}
\end{figure}
\section{\bf Acknowledgments}
I am very grateful to Bo-Sture Skagerstam and Per Elmfors, whom many of the
results  here presented have been obtained in collaboration with.
I also wish to express my appreciation to the organizers of ``Thermo-95'',
particularly professor Gui.
\section{\bf References }
\begin{enumerate}
\item \label{elmforsps94} P.~Elmfors, D.~Persson and B-S.~Skagerstam,
               \prl{71}{1993}{480}, and
               \app{2}{1994}{299}.
\item \label{elmforsps95}  P.~Elmfors, D.~Persson and B-S.~Skagerstam,
	preprint G\"oteborg ITP 95-15, hep-ph 9509418, submitted to
	 {\it Nucl.~Phys.}
\item \label{furry} W.~H.~Furry, \pr{81}{1951}{115}.

\item \label{mak95}  K.~W.~Mak
       \pr{D49}{1994}{6939}
\item \label{weldon89} H.~A.~Weldon,
        \pr{D40}{1989}{2410}.
\item \label{elmfors95} P.~Elmfors, these proceedings.
\item \label{braatenp90} E.~Braaten and R.~D.~Pisarski, \np{B337}{1990}{569}.
\end{enumerate}
\end{document}